\documentclass[twocolumn,superscriptaddress,pra,showpacs]{revtex4}
\usepackage{graphicx}
\usepackage{dsfont}
\usepackage{amsmath}
\usepackage{color}

\newcommand{\trace}{\mathrm{Tr}\,}

\newcommand{\bra}[1]{\langle #1|}
\newcommand{\ket}[1]{|#1\rangle}

\newcommand{\expt}[1]{\langle #1\rangle}

\newcommand{\transpose}{^\mathrm{T}}

\newcommand{\eprint}[1]{\href{http://www.arxiv.org/abs/#1}{\texttt{[#1]}}}

\begin{document}
\title{Phase transitions and localizable entanglement in cluster-state spin chains \\ with Ising couplings and local fields}
\author{Stein Olav Skr{\o}vseth}
\affiliation{Norwegian University of Science and Technology, 7491
Trondheim, Norway}%
\affiliation{School of Physics, The University of Sydney, Sydney,
New South Wales 2006, Australia}
\author{Stephen D. Bartlett}%
\affiliation{School of Physics, The University of Sydney, Sydney,
New South Wales 2006, Australia}
\date{14 August 2009}

\begin{abstract}
We consider a one-dimensional spin chain for which the ground state
is the cluster state, capable of functioning as a quantum
computational wire when subjected to local adaptive measurements of
individual qubits, and investigate the robustness of this property
to local and coupled (Ising-type) perturbations. We investigate the
ground state both by identifying suitable correlation functions as
order parameters, as well as numerically using a variational method
based on matrix product states. We find that the model retains an
infinite localizable entanglement length for Ising and local fields
up to a quantum phase transition, but that the resulting entangled
state is not simply characterized by a Pauli correction based on the
measurement results.
\end{abstract}

\maketitle

\section{Introduction}

Measurement-based quantum computation (MBQC)~\cite{Briegel:2009} has
recently emerged as an alternative model for quantum computation to
the standard circuit model~\cite{Nielsen&Chuang}.  In MBQC,
computation proceeds via local (single-qubit) adaptive measurements
on a fixed fiducial state of a quantum many-body system, for which
the cluster state~\cite{Briegel:2009, Raussendorf:2003p1114} is the
canonical example.

The MBQC model of quantum computation is particularly useful for
investigating the physical requirements for quantum computing, as it
becomes possible to pose questions about universality and
fault-tolerance in terms of properties of quantum
states~\cite{Raussendorf:2005p186, Raussendorf:2007p1547,
Gross:2007, gross:052315, VanDenNest:2007, Brennen:2008,
Doherty:2008, Barrett:2008, GrossFlammia:2008, Bremner:2008}. For
example, the universality of a given quantum state for MBQC may be
determined by assessing the fidelity and range of a universal
quantum gate set, which in turn is quantified by viewing gates as
resource states for gate teleportation~\cite{GottesmanChuang:1999,
Raussendorf:2003p1114} prepared via local
measurements~\cite{Chung:2008}.

However, despite this powerful framework, there are relatively few
known examples of resources states (or distinct classes of states)
that allow for MBQC~\cite{Gross:2007,gross:052315,VanDenNest:2007}.
A promising avenue for identifying properties of states that allow
for MBQC is by investigating ground or low-temperature thermal
states of a coupled quantum many-body
system~\cite{Bartlett:2006p1057,Griffin:2008,Brennen:2008}. In
particular, one can construct model Hamiltonians for which the
ground state is universal for MBQC (say, the cluster state on some
appropriate lattice) and then investigate the robustness of this
property to local perturbations.  Progress has been made in this
direction for thermal or local perturbations of the cluster
state~\cite{Raussendorf:2005p186, Doherty:2008, Barrett:2008}.

In this paper, we investigate a 1D chain of qubits for which the
cluster state is the ground state, and investigate the robustness of
its computational power to both local and coupled (Ising-type)
perturbations.  (For a 1D chain, ``computational power'' is
restricted to single-qubit unitary evolution.)  Consistent with
previous results~\cite{Doherty:2008,Barrett:2008}, we identify a
robust \emph{phase} for which every ground state can allow for
quantum information to be transferred using local measurements.  We
investigate the usefulness for states in this phase to serve as
``quantum computational wires''~\cite{GrossEisert:2008}, which
function as primitives for MBQC.

\subsection{Quantum computational wires}

Within MBQC, the simplest primitive is the ability to move
information (i.e., teleport) along one-dimensional channels, with
single-qubit unitaries determined by the choice of measurements.
States with this property are known as \emph{quantum computational
wires}~\cite{GrossEisert:2008}. A wire in MBQC consists of two
parts: (i) creating a maximally-entangled state between two distant
points via local measurements; and (ii) identifying the correct
``bi-product'' unitary based on the measurement result (either to
implement the identity gate, or some more general single-qubit
unitary gate). The first property is characterized by
\emph{localizable entanglement}~\cite{Popp:2005}, which is the
maximum average entanglement that can be localized on two sites
through local measurements on all other sites.  For systems where
the localizable entanglement falls of exponentially, the
entanglement length $\xi_E$ is defined through
\begin{equation}
  E_L \sim e^{-n/\xi_E}\qquad\text{when } n\gg1\,,
\end{equation}
with $n$ the separation between the sites; states with finite
localizable EL are not directly useable as a quantum computational
wire.  In contrast, systems with infinite localizable EL allow for
teleportation over arbitrary length scales.  One of the
characteristics of the cluster state is a diverging localizable EL.
(We note, however, that diverging EL is in itself not necessary nor
sufficient for the state to be useful for MBQC~\cite{Gross:2007}.)

In general, identifying an optimal measurement basis for localizing
entanglement can be a challenge. For the cluster state on any
lattice, such an optimal measurement strategy is known: measure $Z$
(i.e., measure in the eigenbasis of $\sigma^z$) on all qubits except
those on a line connecting the two desired qubits, thus effectively
making a 1D cluster state, and then measure $X$ on all intermediate
qubits along the remaining line. This measurement sequence will
concentrate a maximally-entangled state on the two remaining qubits.

To transform the resulting maximally-entangled state into a
particular one (say, the two-qubit cluster state), a correction
unitary on one of the qubits based on the measurement results is
then applied.  If these corrections are Pauli operations, or in
general if they close to form a finite subgroup, then the bi-product
operators do not need to be performed but instead one can use a
classical computer to keep track of them.  In general, however, even
if the localizable EL is infinite, the bi-product operators may not
close in this way.  That is, there is a distinction between infinite
localizable EL and the ability to function as a quantum
computational wire.

This paper is structured as follows.  In Sec.~\ref{sec:ClusterHam},
we introduce our Hamiltonian with local and coupled perturbations,
and investigate its phase diagram.  Here, we identify a
\emph{cluster phase} connected to the cluster state (with zero
perturbations).  We consider using correlation functions as for the
cluster state to quantify the identity gate within the cluster phase
in Sec.~\ref{Sec:Exptval}. In Sec.~\ref{sec:Numerics}, we use
numerical methods to assess the localizable entanglement within our
model, and in Sec.~\ref{Sec:semicomputable} we investigate the
usefulness of ground states in the cluster phase to serve as a
quantum computational wire.  In Sec.~\ref{Sec:Disentangle} we turn
our attention to the use of local measurements to \emph{disentangle}
the two halves of the chain, in analogy to the $Z$ measurement in
cluster state MBQC, and finish with conclusions in
Sec.~\ref{Sec:Conclusions}.  The appendix provides details on the
Jordan-Wigner transformation.

\section{Cluster Hamiltonian with anomalous terms}
\label{sec:ClusterHam}

\subsection{The cluster state and cluster Hamiltonian}

Consider a graph (such as a lattice) containing $N$ vertices, with a
qubit placed on each vertex. The cluster state on this graph can be
constructed by first preparing each qubit in the state $\ket +$, the
$+1$ eigenstate of the $\sigma^x$ Pauli spin operator, and then
applying a controlled sign operator $U=\exp(i\pi |1\rangle\langle 1|
\otimes |1\rangle \langle 1|)$ on every pair of qubits on vertices
connected by an edge. The resulting cluster state is compactly
described in terms of stabilizers. That is, at every vertex
$\mu=1,\ldots,N$ one associates an operator
\begin{equation}
  K_\mu=\sigma^x_\mu\prod_{\nu\sim\mu}\sigma^z_\nu\,,
\end{equation}
where $\nu\sim\mu$ indicates all vertices $\nu$ that are connected
to $\mu$.  In a graph of degree $\chi$, this operator acts on
$\chi+1$ vertices. The cluster state is the unique state $\ket C$
that satisfies
\begin{equation}
  K_\mu\ket C=\ket C\,, \qquad \forall\ \mu \,.
\end{equation}

An alternate approach to prepare a cluster state would be to view
the qubits on this graph as an interacting quantum many-body system,
and to find some Hamiltonian $H_C$ for which the cluster state is
the unique ground state. Provided the model has a sufficiently large
gap, preparation of the state would amount to cooling the system
down to (or near) the ground state.  One such Hamiltonian that has
the cluster state as its ground state is (minus) the sum of the
stabilizers at every site $\mu$ \cite{Raussendorf:2005p186},
\begin{equation}
    H_C=-\sum_\mu K_\mu\,.
    \label{CSHamiltonian}
\end{equation}
We have chosen units such that the energy scale is fixed, and the
gap in the model is 2 between the unique ground state and a $N$-fold
degenerate first excited state.  Although the terms $K_\mu$ is this
Hamiltonian are generally many-body, such interactions can occur as
the low-energy behavior of a more natural two-body
Hamiltonian~\cite{Bartlett:2006p1057, Griffin:2008}.  We note that
the cluster state on a line -- the ground state of the Hamiltonian
(\ref{CSHamiltonian}) -- has an infinite localizable EL and can
serve as a quantum computational wire.  This Hamiltonian on a line
can be realized in an optical lattice despite its three-body
interactions \cite{Pachos:2004p1778}.

\subsection{Cluster Hamiltonian with anomalous terms}

Consider the cluster state Hamiltonian (\ref{CSHamiltonian}) in one
dimension with two types of additional terms: local fields and
couplings,
\begin{equation}
    H=H_C-\sum_\mu\vec
    B\cdot\vec\sigma_\mu -J\sum_\mu\sigma^z_\mu\sigma^z_{\mu+1} \,,
    \label{Eq:Hamiltonian}
\end{equation}
with $J\geq 0$.  Unless otherwise specified, we will consider
periodic boundary conditions, such that
$\vec\sigma_{N+1}=\vec\sigma_1$. Throughout the article we will
restrict our attention to $B_y=0$.

(Note that an Ising interaction as in the last term of Eq.
(\ref{Eq:Hamiltonian}) can be used in a time modulated fashion to
construct a cluster state \cite{Raussendorf:2003p1114}. However, we
will study this term only as a constant perturbation of the
Hamiltonian.)

A number of results are known for specific cases involving a single
local term and $J=0$:

\textbf{Local $B_z$ field:} In one dimension, the system is
fundamentally unstable with the addition of a local $z$ field, i.e.,
the EL becomes finite for any non-zero $B_z$.  In two- or
higher-dimensional lattices, however, it can be shown using
techniques from percolation theory that the model exhibits a
transition from a finite region of parameter space $B_z < B_z^{\rm
crit}$ wherein the ground state has infinite localizable EL. Because
the localizable EL is zero in the limit $B_z \to \infty$, there is a
transition in the localizable EL for two- or higher-dimensional
lattices even though the underlying model does not exhibit any
quantum phase transition \cite{Barrett:2008}.

\textbf{Local $B_x$ field:} In one dimension, the system exhibits a
single quantum phase transition at $B_x=1$ separating the ``cluster
phase'' and a separable phase.  Ground states in the cluster phase
are characterized by infinite localizable EL \cite{Pachos:2004p1778,
Doherty:2008}, and the system serves as a quantum computational wire
at all length scales (with precisely the same measurement sequence
and corrections as for the cluster state) albeit with lower fidelity
for all $B_x<1$.  The performance as a quantum computational wire is
therefore a robust property of this system in the presence of a
perturbing $B_x$ field.

With these prior results, we now analyze the full phase space of the
Hamiltonian (\ref{Eq:Hamiltonian}).  Consider the action on this
model of the unitary transformation
$\mathbf{U}=\prod_{\mu\nu}U_{\mu\nu}$ that applies the controlled
phase gate $U_{\mu\nu}$ on all pairs of adjacent qubits. This
unitary maps $H_C$ to
\begin{equation}
  \mathbf UH_C\mathbf U^\dag=-\sum_\mu\sigma_\mu^x\,,
\end{equation}
for which the ground state is separable, $\mathbf U\ket
C=\ket+^{\otimes N}$. In general, the transformation leaves
$\sigma^z$ operators invariant, while $\sigma^x$ maps to
\begin{equation}
  \mathbf U\sigma^x_\mu\mathbf U^\dag=\sigma^x_\mu\prod_{\nu\sim\mu}\sigma^z_{\nu}=K_\mu\,.
\end{equation}

This transformation allows us to determine some properties of the
model (\ref{Eq:Hamiltonian}).  First, consider the model with
$\vec{B}=0$. Under this unitary mapping, the model is dual to the
ordinary transverse-field Ising model, which is completely solved in
1D~\cite{Pfeuty:1970}, and has a single quantum phase transition at
$J=1$. For our model (\ref{Eq:Hamiltonian}) in the large $J$ limit
with $\vec{B}=0$, the ground state approaches a GHZ state
\begin{equation}
  \ket{GHZ}=\frac1{\sqrt2}\left(\ket{00\cdots0}+\ket{11\cdots1}\right)\,.
\end{equation}
We denote the phase for $J>1$ the Ising phase.

From the local $B_z$ results, we do not expect the properties of the
cluster state to survive for any $B_z>0$, and so we now consider the
restricted model with $B_z=0$. This model can be subjected to a
Jordan-Wigner transform as shown in Appendix \ref{Appendix:JW},
which easily allows identification of critical points in the model,
and hence diverging correlation and entanglement lengths. This
restricted model has three clearly identifiable phases: the
\emph{cluster phase} as $B_x,J\to0$, an \emph{Ising phase} as
$J\gg\max(1,B_x)$ and a \emph{separable phase} as $B_x\gg\max(1,J)$.
The model is unitarily dual under the transformation $\mathbf U$.
Thus, for any critical point at $(J,B_x)$, there is another critical
point at $(\frac J{B_x},\frac1{B_x})$.  Thus, considering the
critical line $\mathcal C_C$ connecting the known critical points at
$(0,1)$ and $(1,0)$, this duality reveals another critical line
$\mathcal C_G$ from $(0,1)$ to $(\infty,\infty)$.  Parameterizing
the lines with a parameter $\tau\in[0,1]$, we have
\begin{equation}
  \mathcal C_C=(\tau,f(\tau))\qquad \mathcal C_G=\left(\frac
  \tau{f(\tau)},\frac1{f(\tau)}\right)\,,
\end{equation}
for some unknown function $f(\tau)$, for which $f(0)=1$ and
$f(1)=0$. In the limit $B_x,J\gg1$, the cluster term in the
Hamiltonian becomes unimportant, and the model is a simple Ising
model for which the critical point must be at $B_x=J$. Under the
above parameterization we must have, assuming $f(\tau)$ is
continuous, and invoking l'H\^{o}pital's rule,
\begin{equation}
  \lim_{\tau\to1}\frac\tau{f(\tau)}-\frac1{f(\tau)}
  =\lim_{\tau\to1}\frac{\tau-1}{f(\tau)}=\lim_{\tau\to1}\frac1{df/d\tau}=0\,,
\end{equation}
which implies $\frac{df}{d\tau}=\infty$. Hence, the critical line
$\mathcal C_C$ must be convex (at least close to $J=1$). The
numerical results indicate that this convergence is very slow, and
to a close approximation we have $f(\tau)=1-\tau$, and hence
$\mathcal C_G=(\tau',1+\tau')$, $\tau'\in[0,\infty]$. The three
phases are separated by Ising type critical lines corresponding to a
conformal field theory with central charge $c=1/2$, and a
corresponding entanglement signature \cite{Skrovseth:2005}.

\begin{figure}
    \includegraphics[width=\columnwidth]{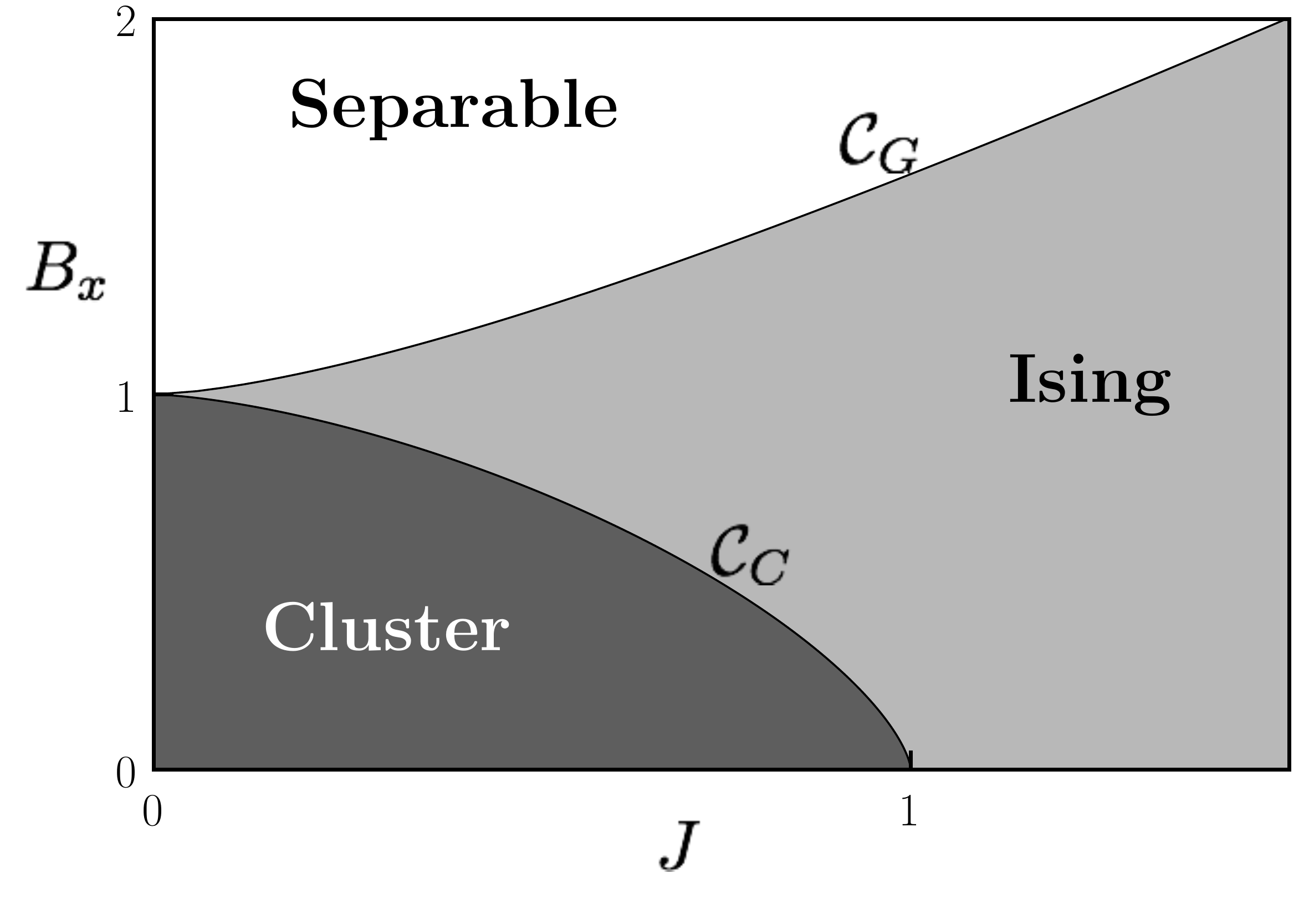}
    \caption{A sketch of the phase diagram for the Hamiltonian in
    Eq.~(\ref{Eq:Hamiltonian}) with $B_z=0$. The lines $\mathcal C_G$
    and $\mathcal C_C$ constitute quantum phase transitions between
    the three indicated phases.  (All quantities shown in figures in this paper,
    including this one, are dimensionless.)}
    \label{Fig:phasediag}
\end{figure}

In the phase diagram of Fig.~\ref{Fig:phasediag} for $B_z=0$, at the
origin ($J=B_x=0$) the ground state is the cluster state.  From the
results of~\cite{Pachos:2004p1778, Doherty:2008}, we know that
ground states on the line $J=0$, $0\leq B_x < 1$ have infinite
localizable EL, and in fact allow for long-ranged single-qubit gates
using the same measurement sequence and Pauli corrections as for the
cluster state. We now turn our attention to the question:  is any
ground state within the cluster phase of Fig.~\ref{Fig:phasediag}
also useful as such a quantum computational wire?

\section{Correlation functions for localizable entanglement and quantum computational wires}
\label{Sec:Exptval}

In special cases, the localizable entanglement of a state can be
characterized (or, at least, lower bounded) by a correlation
function.  For example, with certain measurement sequences, one can
obtain post-measurement quantities from pre-measurement expectation
values to quantify the localizable
entanglement~\cite{VerstraetePoppCirac:2004,Popp:2005,Venuti:2005}
or the performance of quantum gates in MBQC~\cite{Chung:2008,
Doherty:2008}.  That is, the expectation values of string-like
operators can serve as order parameters to identify the hidden
correlations corresponding to localizable entanglement and the
ability to function as a quantum computational wire.

\subsection{Localizable entanglement for $B_z=0$}

For $B_z=0$, the symmetries of our model allow us to determine the
optimal measurement sequence and to derive an expression for the
localizable
entanglement~\cite{VerstraetePoppCirac:2004,Popp:2005,Venuti:2005}.
Our model (\ref{Eq:Hamiltonian}) with $B_z=0$ is invariant under
rotations of all spins by $\pi$ about the $x$ axis, transforming
$\sigma^z \rightarrow - \sigma^z$.  It is not invariant under
similar rotations about the $y$ and $z$ axes; therefore, the optimal
measurements for localizing entanglement on a finite chain between
the end qubits $1$ and $N$ are measurements in the
$X$-basis~\cite{VerstraetePoppCirac:2004,Popp:2005,Venuti:2005}.
With this basis, the localizable entanglement for this finite chain
is given by the string correlation function
\begin{equation}
  E_L = \langle
  \sigma^y_1\Bigl({\textstyle\prod_{j=2}^{N-1}}\sigma^x_{j}\Bigr)\sigma^y_{N}\rangle
  \,.
  \label{eq:LEstringcorrelation}
\end{equation}
(We note that, for a suitable choice of boundary conditions on this
finite chain, we could ensure that the ground state is a $+1$
eigenstate of $P^x = \prod_{j=1}^{N} \sigma^x_j$ and simplify this
correlation function further.  However, as we will be interested in
open chains or periodic boundary conditions, we will not do so.)

This correlation function quantifies the localizable entanglement,
but does not completely characterize the form of the resulting
maximally entangled state. We now turn to a complete set of
correlation functions for this, which quantify the state's use as a
quantum computational wire.

\subsection{Correlation functions for quantum computational wires}

In this section, we briefly summarize the main result of
Ref.~\cite{Chung:2008}. Consider a lattice of qubits prepared in
some initial pure state $|\psi_0\rangle$. Singling out two qubits,
$a$ and $b$, we consider a measurement sequence on the remaining
qubits in the lattice that localizes entanglement on $a$ and $b$.
Let $m$ label the measurement outcomes, and $P_m$ be the
corresponding projector. Following the measurements, a correction
unitary $U_m$ conditional on $m$ is applied to qubit $b$. Averaged
over all possible measurement outcomes, the resulting two-qubit
state is
\begin{equation}
  \rho_{ab} = {\textstyle \sum_m} U_m P_m |\psi_0\rangle \langle\psi_0 | P_m U_m^\dagger\,.
\end{equation}
Equivalently, we can characterize this final state using the set of
expectation values of bipartite Pauli operators $\sigma^i_a \otimes
\sigma^j_b$, $i,j=I,x,y,z$ on qubits $a$ and $b$, as
\begin{align}
  \langle \sigma^i_a \sigma^j_b \rangle_{\mathcal P}
  &=  {\textstyle \sum_m} \langle\psi_0 | P_m U_m^\dagger \sigma^i_a \sigma^j_b U_m P_m |\psi_0\rangle \nonumber \\
  &=  {\textstyle \sum_m} \langle\psi_0 | P_m \sigma^i_a (\sigma^j_b)_mP_m |\psi_0\rangle\,,
\end{align}
where $(\sigma^j_b)_m = U_m^\dagger \sigma^j_b U_m$, and where
$\langle \cdot\rangle_{\mathcal P}$ denotes the expectation value in
the final post-measurement two-qubit state. The set of such
correlation functions for all pairs of Pauli operators will
completely specify the two-qubit state.

We now restrict our attention to measurement sequences for which
there exists a string of operators $S$ acting on some set of the
measured qubits which is \emph{independent} of the measurement
outcomes, and an operator $\tau^j_b$ on $b$ that is also independent
of $m$, such that
\begin{equation}
  \label{eq:CommutationIdentity}
  \sigma^i_a (\sigma^j_b)_m P_m = P_m \sigma^i_a S \tau^j_b\,.
\end{equation}
For example, in the cluster state model of
MBQC~\cite{Raussendorf:2003p1114}, a universal gate set is known
that satisfies this property~\cite{Chung:2008}. For such measurement
sequences, using the projector properties $P_m^2 = P_m$ and $\sum_m
P_m = I$ gives
\begin{equation}
  \label{eq:CorrFunction}
  \langle \sigma^i_a \sigma^j_b \rangle_{\mathcal P} =  \langle\psi_0|\sigma^i_aS\tau^j_b|\psi_0\rangle\,.
\end{equation}
Thus we can relate the two-qubit state prepared \emph{after} the
sequence of measurements to a correlation function of the original
state $|\psi_0\rangle$ \emph{prior} to measurements.  That is, the
correlation functions characterize the \emph{post-}measurement
two-qubit state using expectation values of strings of operators on
the \emph{pre-}measurement state.

It is critical to this development that one can identify such a
string of operators $S$.  The measurement sequence for localizing
entanglement in the cluster state provides the canonical example.
For a $N$-qubit chain with $N$ even, using the same measurement
sequence as if localizing entanglement in the cluster state
$|C\rangle$ (measure $Z$ on qubits 1 and $N$, and $X$ on all qubits
$3\cdots N-2$) with the same Pauli corrections as for the cluster
state, the averaged state on qubits $a=2$ and $b=N-1$ afterwards has
the following non-zero correlation functions:
\begin{subequations}
\label{ZXXZexptvalues}
\begin{align}
    \expt{\sigma^z_a\sigma^x_b}_{\mathcal P}
    &=\langle C|\sigma^z_2\Bigl({\textstyle\prod_{j=1}^{N/2-1}}\sigma^x_{2j+1}\Bigr)\sigma^z_N|C\rangle \label{ZXexptvalue}\\
    \expt{\sigma^x_a\sigma^z_b}_{\mathcal P}
    &=\langle C|\sigma^z_1\Bigl({\textstyle\prod_{j=1}^{N/2-1}}\sigma^x_{2j}\Bigr)\sigma^z_{N-1}|C\rangle \label{XZexptvalue}\\
    \expt{\sigma^y_a\sigma^y_b}_{\mathcal P}
    &=\langle C|\sigma^z_1\Bigl({\textstyle\prod_{j=2}^{N-1}}\sigma^x_{j}\Bigr)\sigma^z_{N}|C\rangle \label{YYexptvalue}
\end{align}
\end{subequations}
Note that a two qubit cluster state has expectation values
$\expt{\sigma^x_a\sigma^z_b}=\expt{\sigma^z_a\sigma^x_b}=\expt{\sigma^y_a\sigma^y_b}=1$,
which means that the above expectation values are maximal for the
cluster state.

\subsection{Correlation functions for $B_z=0$}

\begin{figure}
    \includegraphics[width=\columnwidth]{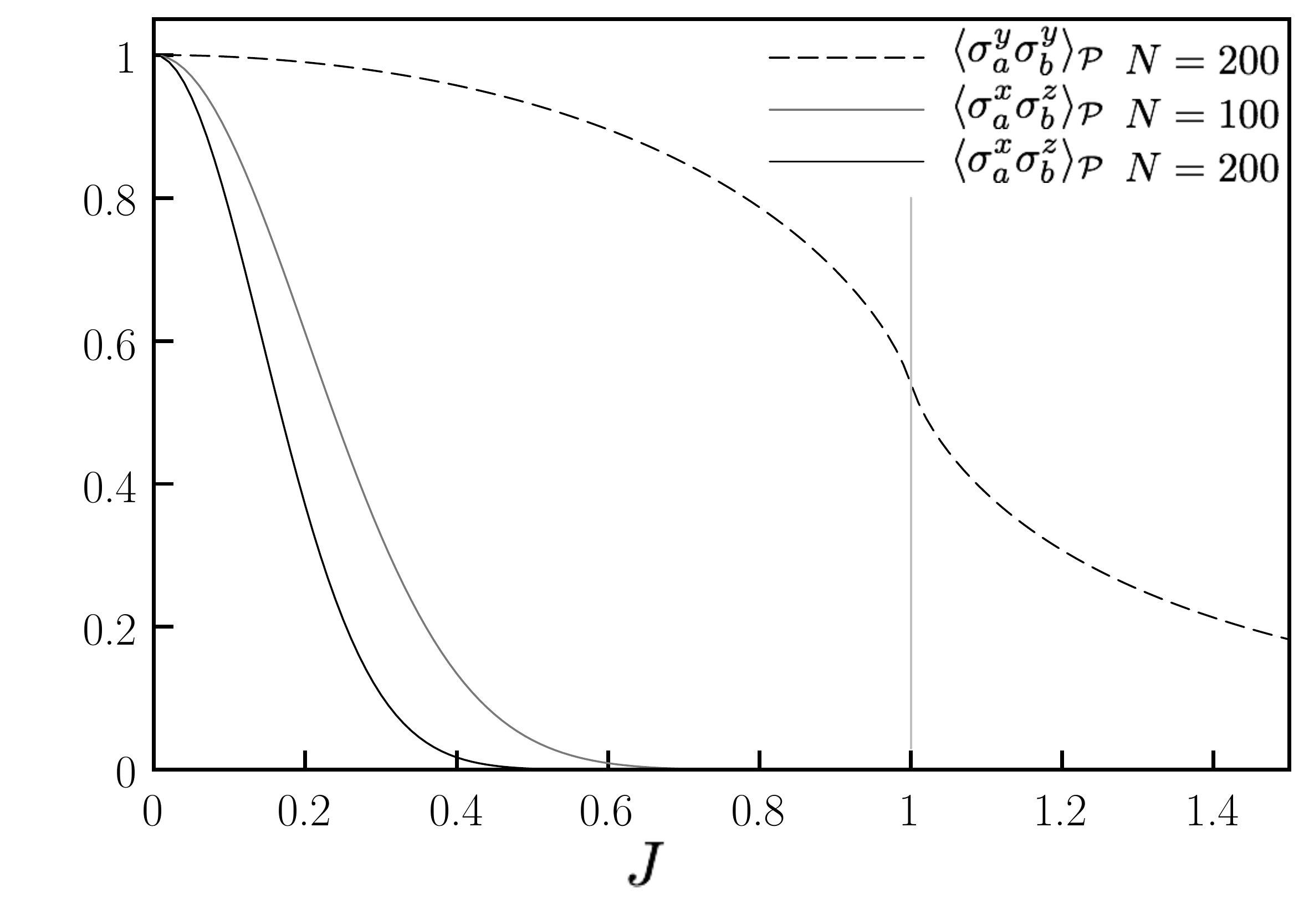}
    \caption{Expectation values $\expt{\sigma^x_a\sigma^z_b}_{\mathcal P}$ and
    $\expt{\sigma^y_a\sigma^y_b}_{\mathcal P}$ for $B_x=B_z=0$. The former is shown for
    $N=100,200$ while $\expt{\sigma^y_a\sigma^y_b}_{\mathcal P}$ is shown for $N=200$
    only, as this is not size dependent, and indistinguishable from
    $N=100$.  The expectation value $\expt{\sigma^z_a\sigma^x_b}_{\mathcal P}$ is identical
    to $\expt{\sigma^x_a\sigma^z_b}_{\mathcal P}$. }
    \label{Fig:Orderparam}
\end{figure}

For our model (\ref{Eq:Hamiltonian}) with $B_z=0$, the correlation
functions of Eq.~(\ref{ZXXZexptvalues}) can be calculated
analytically using the exact solution obtained via the Jordan-Wigner
transformation.  In the cluster phase, of the three correlation
functions given in Eq.~(\ref{ZXXZexptvalues}), we find that only the
expectation value $\expt{\sigma^y_a\sigma^y_b}_{\mathcal P}$ of
Eq.~(\ref{YYexptvalue}) remains long-ranged in the thermodynamic
limit.  This expectation value takes the general form of a
transverse-field Ising order parameter.  The other expectation
values, $\expt{\sigma^z_a\sigma^x_b}_{\mathcal P}$ and
$\expt{\sigma^x_a\sigma^z_b}_{\mathcal P}$, remain nonzero only for
$J=0$. (As shown in~\cite{Doherty:2008}, for $J=0$ they remain
long-ranged for $B_x<1$.) Fig.~\ref{Fig:Orderparam} shows the
behaviour of these expectation values for the ground state of the
model with $\vec{B}=0$.

The long-ranged behaviour of $\expt{\sigma^y_a\sigma^y_b}_{\mathcal
P}$ in the cluster phase serves as a useful order parameter, and is
related to the expression for the finite-chain localizable
entanglement of Eq.~(\ref{eq:LEstringcorrelation}).  However, the
fact that the other correlation functions of
Eq.~(\ref{ZXXZexptvalues}) are zero for $J>0$ shows that the
standard cluster-state Pauli corrections based on the measurement
results are not suitable, i.e., the resulting bipartite state is
\emph{not} the two-qubit cluster state.  We now turn to numerical
methods to investigate the exact form of the resulting bipartite
entangled state in the cluster phase.

\section{Numerical calculations using matrix product states}
\label{sec:Numerics}

Our analytic approach to investigating the cluster phase has
identified an appropriate order parameter for the localizable
entanglement, realized as the expectation value of a string
operator, but has not revealed the form of the resulting bipartite
entangled state necessary for use as a quantum computational wire.
We now use numerical methods based on Matrix Product States to
investigate the localizable EL in this phase.

\subsection{Matrix Product States}

Matrix Product States (MPS) have emerged as the natural language to
describe 1D systems with limited long-ranged entanglement, i.e., off
criticality \cite{Tagliacozzo:2008}. (There are examples of MPS
critical points where the states subscribe to an exact MPS
representation \cite{Wolf:2006}, but general quantum critical points
will be associated with a diverging entanglement.) The
well-established density matrix renormalization group (DMRG)
scheme~\cite{White:1992} has subsequently been shown to be an
iterative minimization over MPS. Such states are also well suited to
describe local measurements, and hence naturally suited to cluster
states and measurement-based quantum computation in
general~\cite{Gross:2007,gross:052315}.  Even when a full analytic
solution is available through the Jordan-Wigner transformation, we
still find that an MPS representation is more useful for this
reason.

Given a sufficiently-large bond dimension $D$, any state $\ket\psi$
can be written as a MPS of the form
\begin{equation}
    \ket\psi=\sum_{s_1\cdots s_N=1}^d\trace\left(A^{s_1}_1\cdots A^{s_N}_N\right)\ket{s_1\cdots
    s_N}\,,
\end{equation}
where $d$ is the local Hilbert space dimension (e.g., $d=2$ for
qubits). Each site $\mu$ is associated with $d$ different $D\times
D$ matrices $A^s_\mu$, $s=1,\ldots,d$, which are normalized to
either of
\begin{equation}
  \sum_{s=1}^d (A^s_\mu)^\dag A^s_\mu=\mathds 1\,,\qquad\text{or}\qquad \sum_{s=1}^d
  A^s_\mu(A^s_\mu)^\dag=\mathds 1\,,
\end{equation}
for all $\mu$. For a translationally invariant state, we have
$A^s_\mu \equiv A^s$ independent of $\mu$.

We now consider MPS descriptions of the ground state of our
Hamiltonian (\ref{Eq:Hamiltonian}) in one dimension.  The cluster
state is described by the simplest nontrivial case, with $D=2$ and
\begin{equation}
  A^0=\begin{pmatrix}1&1\\0&0\end{pmatrix}\,,\quad
  A^1=\begin{pmatrix}0&0\\1&-1\end{pmatrix}\,.
\end{equation}
With $J=B_x=0$ but $B_z>0$, one still has a $D=2$ representation.
Any MPS representation can be obtained if one knows a sequential
procedure to produce the state such that two neighbouring states are
a result of operation by a two-qubit operator $\bar U$ when all
qubits initially are in the state $\ket{0}$
\cite{Perez-Garcia:2006}.  In the cluster state $\bar
U=U(H\otimes\mathds 1)$ where $H$ is the Hadamard transform and $U$
is the controlled phase gate.  For the $B_z>0$ case, the controlled phase gate
commutes with the extra term in the Hamiltonian, so the resulting
ground state can be written as $\ket{C_z}=U\ket\theta^{\otimes N}$
where $\ket\theta=\cos\theta\ket++\sin\theta\ket-$ and
\begin{equation}
  \tan\theta=\frac{\sqrt{B_z^2+1}-1}{B_z}\,.
\end{equation}
Hence, to obtain the MPS one can use the same sequential generation scheme, replacing the Hadamard by
\[H_\theta=\begin{pmatrix}\eta_+&\eta_-\\\eta_-&-\eta_+\end{pmatrix}\]
with $\eta_\pm=(\cos(\theta)\pm\sin(\theta))/\sqrt 2$, which results in a MPS representation
\begin{equation}
  A^0=\begin{pmatrix}\eta_+&\eta_-\\0&0\end{pmatrix}\,,\quad
  A^1=\begin{pmatrix}0&0\\\eta_+&-\eta_-\end{pmatrix}\,.
  \label{eq:TiltedMPS}
\end{equation}
All other points in the phase diagram have only approximate
solutions for any $D<2^{N/2}$.

Given a measurement sequence over the qubits $\mathcal R$ with
outcomes $s_i, i\in\mathcal R$, the resulting state after
measurement with the measured qubits traced out is
\begin{equation}
  \ket{\psi'}=\sum_{s_\iota}\trace\left(A_1^{s_1}\cdots
  A_N^{s_N}\right)\ket{\{s_\iota\}}\,,
\end{equation}
where the sum is over all $\iota\not\in\mathcal R$.  A general
measurement on qubits is given by a direction on the Bloch sphere.
As we have restricted our attention to $B_y=0$, the ground state of
our Hamiltonian will always have real coefficients, and we can
restrict our measurements to directions in the $x-z$ plane, given by
a direction $\xi$.  We define $\xi=0~(\pi/4)$ to correspond to an
$Z~(X)$ measurement.

An optimal measurement basis which localizes the maximum
entanglement is given by the angle that maximizes $\sum_i|\det A_i|$
\cite{Verstraete:2004p087201}. Applying this result to states of the
form of Eq.~(\ref{eq:TiltedMPS}), the MPS representation in a tilted
basis is
\begin{equation}
  A_0^{(\xi)}=\cos\xi A_0+\sin\xi A_1\,,\quad
  A_1^{(\xi)}=-\sin\xi A_0+\cos\xi A_1\,,
\end{equation}
and we get
\begin{equation}
  \sum_i|\det
  A_i^{(\xi)}|=|2\sin(2\xi)\eta_+\eta_-|=|\sin(2\xi)\cos(2\theta)|\,.
\end{equation}
Thus, an optimal measurement basis for localizing entanglement in
this state is for $\xi=\pi/4$, i.e., the $X$ basis, independent of
the magnetic field $B_z$.

\subsection{Variational method}

The problem of finding a MPS representation for the ground state of
a given Hamiltonian is an NP-complete problem \cite{Schuch:2008}.
However, variation methods such as DMRG work well in practice for a
wide variety of Hamiltonians. In addition, the MPS representation
makes it easy to compute expectation values, entanglement, and other
physical quantities, and make comparisons to known values where such
exist. The numerical successive minimization is described in other
works \cite{Verstraete:2004,Popp:2005,Perez-Garcia:2006}, but a
short description of what we denote the Variational Matrix Product
State (VMPS) procedure follows. Starting with a random MPS
representation and corresponding energy, one fixes all matrices
except one, and minimizes the energy with respect to the single
matrix. This minimization amounts to a simple generalized eigenvalue
problem. After normalization of the matrix, one moves to the
neighboring site, and then repeatedly sweeps the lattice back and
forth until convergence is reached. We stop the iteration after a
sufficient number of full sweeps plus half way back in the lattice
where we pick out the two matrices needed, and use these as a
representation for a large chain with periodic boundary conditions.
Thus we avoid boundary effects, and we get a convergent
representation.

The performance of the method is naturally dependent on the initial
choice of representation, and it is useful to repeat the procedure
with different choices to find a consistent solution. In general,
energy and local expectation values converge quickly and
independently of the initial state to the same value. However, even
very subtle differences in these quantities can amount to a large
differences in terms of entanglement quantities.

We can assess the accuracy of our VMPS for $B_z=0$, by comparison
with the analytic solution available via the Jordan-Wigner
transform.  With the exact solution, we can easily compute any
$n$-partite reduced entropy $S_n=S(\rho_n)$, where
$\rho_n=\trace_{n+1\cdots N}\ket\psi\bra\psi$ is the reduced state
of $n$ neighboring qubits for a system in the pure state $\ket\psi$,
and $S(\rho)$ is the von Neumann entropy. The single-qubit reduced
entropy $S_1$ will not distinguish the cluster and Ising phases, as
both will have $S_1=1$, so we use the bipartite reduced entropy
$S_2$ as an indicator. This entropy separates the three conventional
phases of $S_2=2$ (cluster), $S_2=1$ (Ising) and $S_2=0$
(separable). A state that is very close to the true ground state in
terms of the energy and expectation values can still be very
different in terms of $S_2$, so it is necessary to run the iteration
several times with different initial condition to obtain a state
that reflects the ground state in this respect.

\begin{figure}
    \includegraphics[width=\columnwidth]{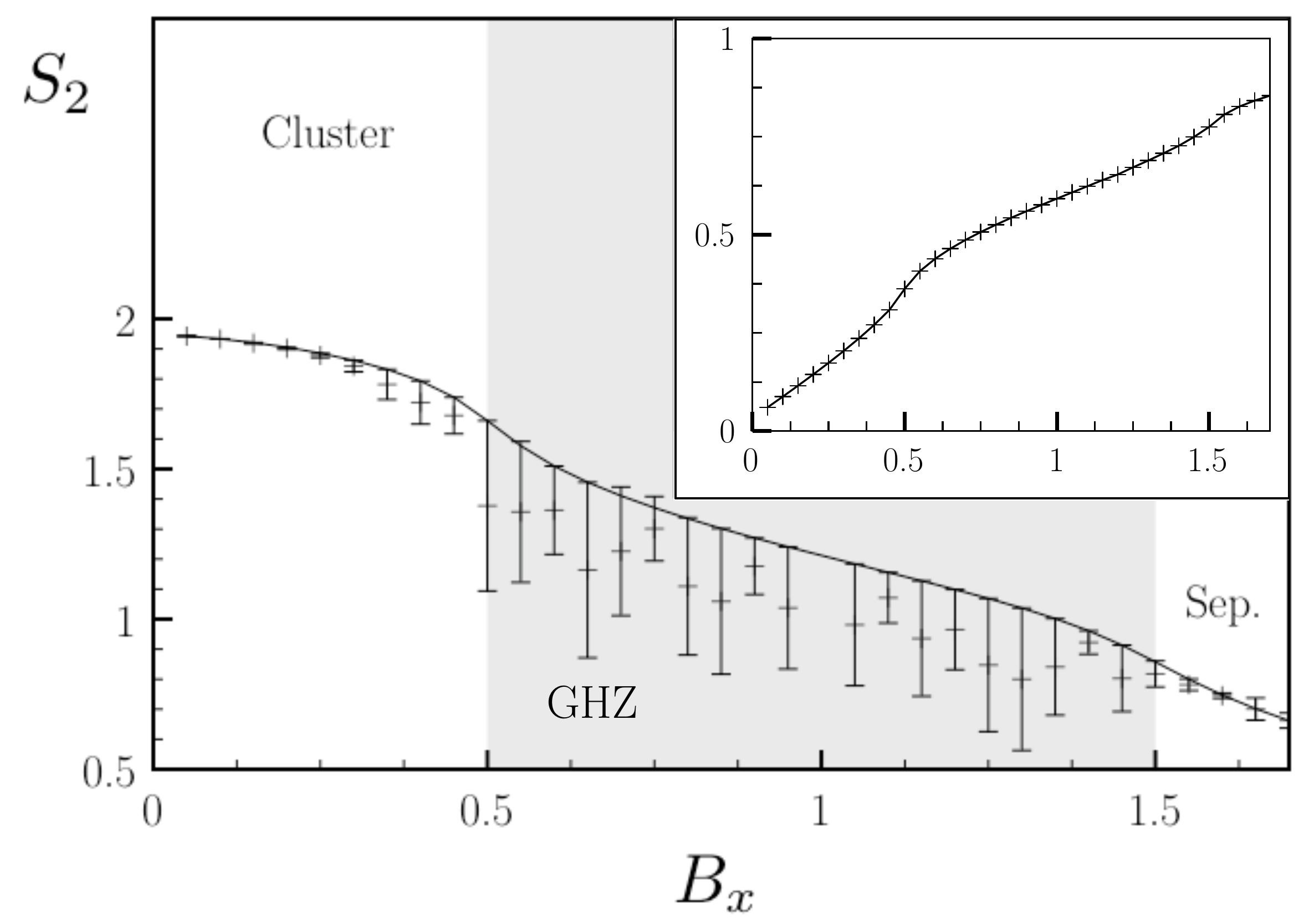}
    \caption{The bipartite
    reduced entropy $S_2$ for $J=0.5, B_z=0$ computed exactly through the
    Jordan-Wigner transform (full line) and through 40 runs of VMPS with 6
    sweeps. Note that the cluster and separable phases compute the
    entanglement properties very precisely, especially away from the phase
    transitions, while in the Ising phase the results are much less reliable.
    Also, the algorithm inevitably underestimates the true value of $S_2$.
    The inset shows the local expectation value $\expt{\sigma^x_\mu}$,
    where error bars are too small to be evident. (N=200, D=8)}
    \label{Fig:twotangle}
\end{figure}

We consider the line $B_z=0, J=0.5$, where we expect two Ising class
phase transitions at $B_x\approx0.5, 1.5$. The bipartite reduced
entropy for a number of runs with random initial conditions are
shown in Fig. \ref{Fig:twotangle}, where it is clear how the
entanglement properties fail for ground states in the Ising phase
(which are GHZ-like), while being relatively good for the other two
phases. The local expectation value $\expt{\sigma^x_\mu}$ is however
well represented for all phases.
These results also show that $S_2$ is upper bounded by the true
value, and the best representation can be chosen as those with the
highest $S_2$ even when the exact value is not known.

\subsection{Numerical estimation of localizable entanglement}

We now seek find an approximate MPS representation for the ground
state of our model and with it, to compute the localizable
entanglement. The latter step can be accomplished by a Monte-Carlo
scheme \cite{Popp:2005} to sample over entanglement by a weighted
random walk in the probability space. Given that two substantial
numerical steps are needed in this method to obtain the localizable
entanglement, and in particular as a tendency of the MPS
construction to occasionally yield incorrect representations of the
ground state, the values obtained has an intrinsic error that we
have indicated wherever applicable. Specifically, in the Ising
phase, the method is unstable and picks out a ground state close to
a product state rather than the GHZ state. However, we are
interested in what happens in the cluster phase, and that problem is
therefore not substantial. Importantly, the technique seems to be
particularly well behaved in the cluster phase.

We use periodic boundary conditions (PBC) throughout, as this makes
the computation much easier and, due to the resulting
translation-invariance, reduces storage to only two matrices as
opposed to $2N$ matrices in the open boundary condition case. As we
show in Sec.~\ref{Sec:Disentangle}, the $Z$ measurements do not
perfectly disentangle the chain for $J>0$, and one might worry that
the results are dependent on the boundary conditions as there are
two directions in the lattice that the entanglement might propagate.
However, our data show that this does not happen. Specifically, one
can do $n_z$ subsequent $Z$ measurements, and for $n_z>1$, the
resulting values for the entanglement is independent of $n_z$ for
all $J$. Hence we are justified to use PBC, and the error in doing
so is much smaller than the statistical noise of our methods, and we
use $n_z=2$ in the following.

To estimate the localizable entanglement we must choose a specific
measurement protocol, and the results of Sec.~\ref{Sec:Exptval}
reveal that the optimal basis for measuring the intermediate qubits
is the $X$ basis.  We measure the qubits beyond the endpoints in the
$Z$ basis as for the cluster state.

Our numerics confirm our expectation that the localizable EL becomes
finite for any $B_z>0$ (as can be shown analytically at $J=B_x=0$
using the MPS representation of Eq.~\ref{eq:TiltedMPS}). For the
remainder of this paper, we restrict to $B_z=0$. We have computed
the localizable entanglement along two lines in the phase diagram of
Fig. \ref{Fig:phasediag}: (i) $B_x=B_z=0$ and (ii) $B_x=.5$ and
$B_z=0$. The (i) line will have an Ising class quantum phase
transitions separating the cluster and Ising phases, while the (ii)
line will have a phase transition at roughly $J=0.5$ between the
same phases.

\begin{figure}
    \includegraphics[width=\columnwidth]{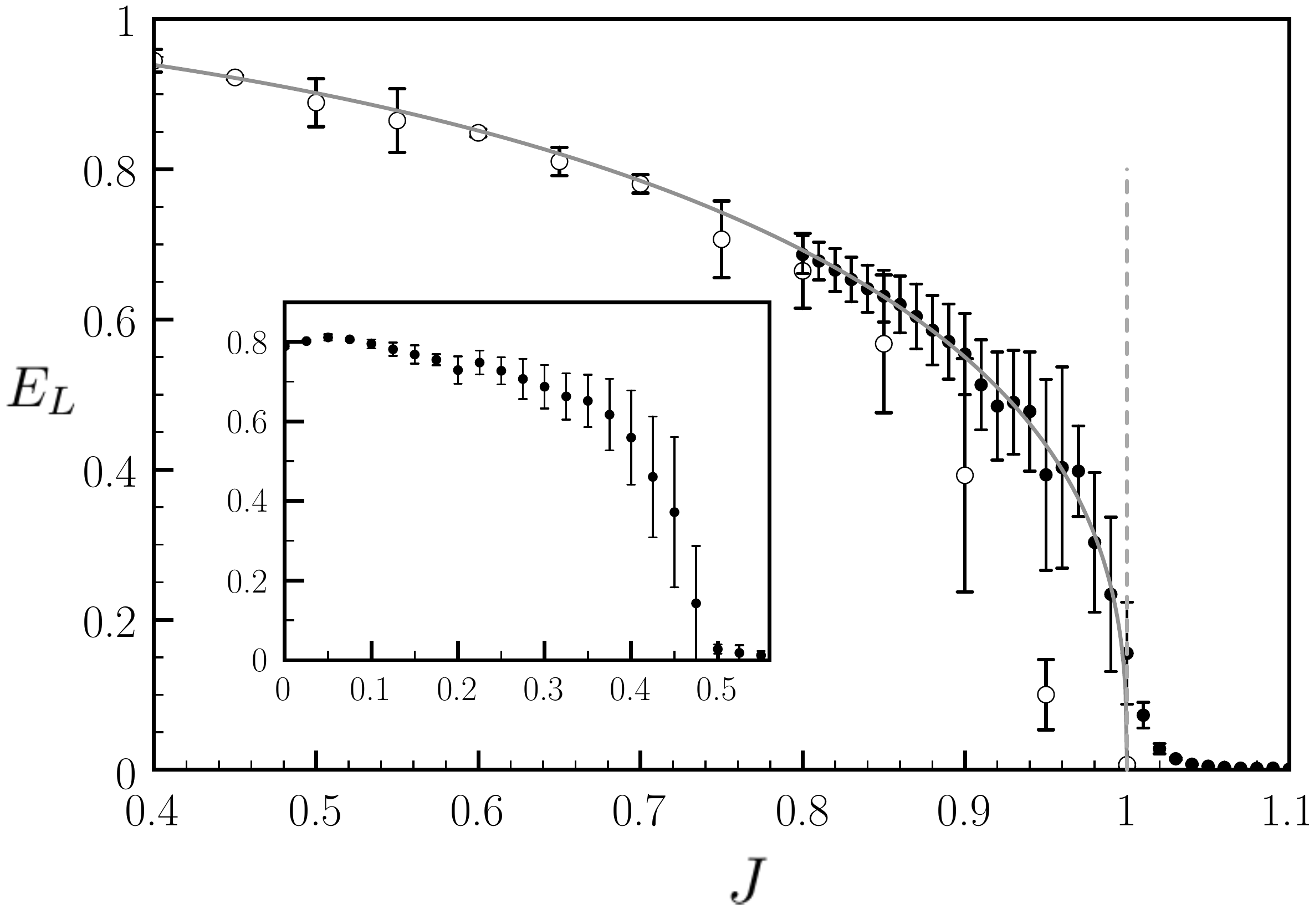}
    \caption{The localizable entanglement for a $N=60$ model with $\vec{B}=0$.
    MPSs used are $D=4$ (open circles) and $D=8$ (solid circles). The gray line
    is a guide to the eye, following the function $(1-J^2)^{0.36}$, which is
    consistent with the expected behaviour of an second order phase transition
    order parameter. The error bars indicate the sensitivity of the initial
    conditions in the VMPS scheme. The inset shows a similar plot for $B_x=.5$
    and $D=8$. Note that the performance of the VMPS is
    noticeably worse as $J$ approaches the phase transition, though the data
    are still conclusive.}
    \label{Fig:ClusterIsing}
\end{figure}

For the model with $\vec{B}=0$, the data clearly shows an infinite
EL for $J<1$, and a significant localizable entanglement, as shown
in Fig. \ref{Fig:ClusterIsing}.  Similar results are found along the
line (ii), with $B_x=0.5$.

The sensitivity on the initial conditions in the VMPS algorithm is
far larger than the numerical errors of the Monte Carlo procedure.
However, the latter is well suited to detect polynomial versus
exponential decay, and performs consistently in this respect for
different initial conditions even though the actual entanglement
values may have substantial variation. Hence, we are able to
separate an infinite EL phase (the cluster phase) and a finite EL
phase.

\section{Characterizing entanglement in the cluster phase}
\label{Sec:semicomputable}

We have shown both analytically and numerically that the localizable
EL is diverging in the cluster phase, and the numerics also provide
a mechanism to analyze the form of the resulting two-qubit entangled
state.  For $J>0$, the resulting two-qubit state is not identical to
the form of the two-qubit cluster state even after the Pauli
corrections. In this section, we explore the form of the
post-measurement state and its dependence on the measurement
results.  We focus specifically on $\vec{B}=0$, but our results
extend directly to $B_x>0$.

As we are performing the identical measurement sequence for
localizing entanglement in the cluster state, it is natural to apply
the same Pauli corrections to the final two-qubit state depending on
the parity of the even and odd $X$-measurements, as well as the
boundary $Z$-measurements.  The correlation functions
(\ref{ZXXZexptvalues}) which incorporate these corrections show that
the resulting state does not take a fixed form.  With the above
Pauli corrections, the sampled states in the Monte Carlo analysis
with near-maximal entanglement are characterized by
$\expt{\sigma^y_a \sigma^y_b}_\mathcal{P}=1$, as expected from our
analysis using the Jordan-Wigner transformation, as well as the
relations
\begin{equation}
    \expt{\sigma^z_a \sigma^x_b}_\mathcal{P}=\expt{\sigma^x_a \sigma^z_b}_\mathcal{P},\
    \expt{\sigma^x_a \sigma^x_b}_\mathcal{P}=-\expt{\sigma^z_a \sigma^z_b}_\mathcal{P},
    \label{zxxz_cond}
\end{equation}
and
\begin{equation}
  \expt{\sigma^z_a \sigma^x_b}_\mathcal{P}^2+\expt{\sigma^x_a \sigma^x_b}_\mathcal{P}^2=1\,.
\end{equation}
(Sampled states that are not near-maximally entangled do not fall
into this class.)  The conditions (\ref{zxxz_cond}) imply that the
maximally entangled state are described by
\begin{equation}
    \ket{\Phi}=\cos\Phi\ket{C_{00}}+\sin\Phi\ket{C_{11}}\,,
\end{equation}
where we have defined
\begin{equation}
  \ket{C_{ij}}=\left(\sigma^x_b\right)^i\left(\sigma^z_b\right)^j\ket
  C\qquad i,j=0,1\,.
\end{equation}
This form of the states is consistent with the long-ranged behaviour
of $\expt{\sigma^y_a \sigma^y_b}_\mathcal{P}$, but reveals an
additional $y$-rotation by an angle $\Phi$.

Investigating the distribution of angles $\Phi$ for states sampled
from the Monte Carlo analysis, we find that these angles do not take
fixed multiples of $\pi/2$.  This state may be corrected into the
cluster state by a $y$-rotation of qubit $b$ by an angle $-\Phi$,
but not by a Pauli correction. The angle $\Phi$ is a function of the
measurement results, but we have been unable to determine this
dependence.

We model $\Phi$ as sampled from a probability distribution $\mathcal
D_J(\Phi)$, satisfying $\mathcal D_0(\Phi)=\delta(\Phi)$.  We now
show that this distribution does not possess any bias away from
zero; if that were the case, one could improve the fidelity of the
post-measurement state with the two-qubit cluster state by
performing a correcting rotation.  For a given $J$, the expectation
value $\expt{\sigma^x_a \sigma^z_b}_\mathcal{P}$ will take the form
\begin{equation}
  \expt{\sigma^x_a \sigma^z_b}_\mathcal{P} = A(J)\cos(\xi(J))\,,
\end{equation}
where $A(0)=1$, $\xi(0)=0$. The phase $\xi(J)$ thereby determines
the bias, and the amplitude $A$ gives the magnitude of the
expectation value.

\begin{figure}
    \includegraphics[width=\columnwidth]{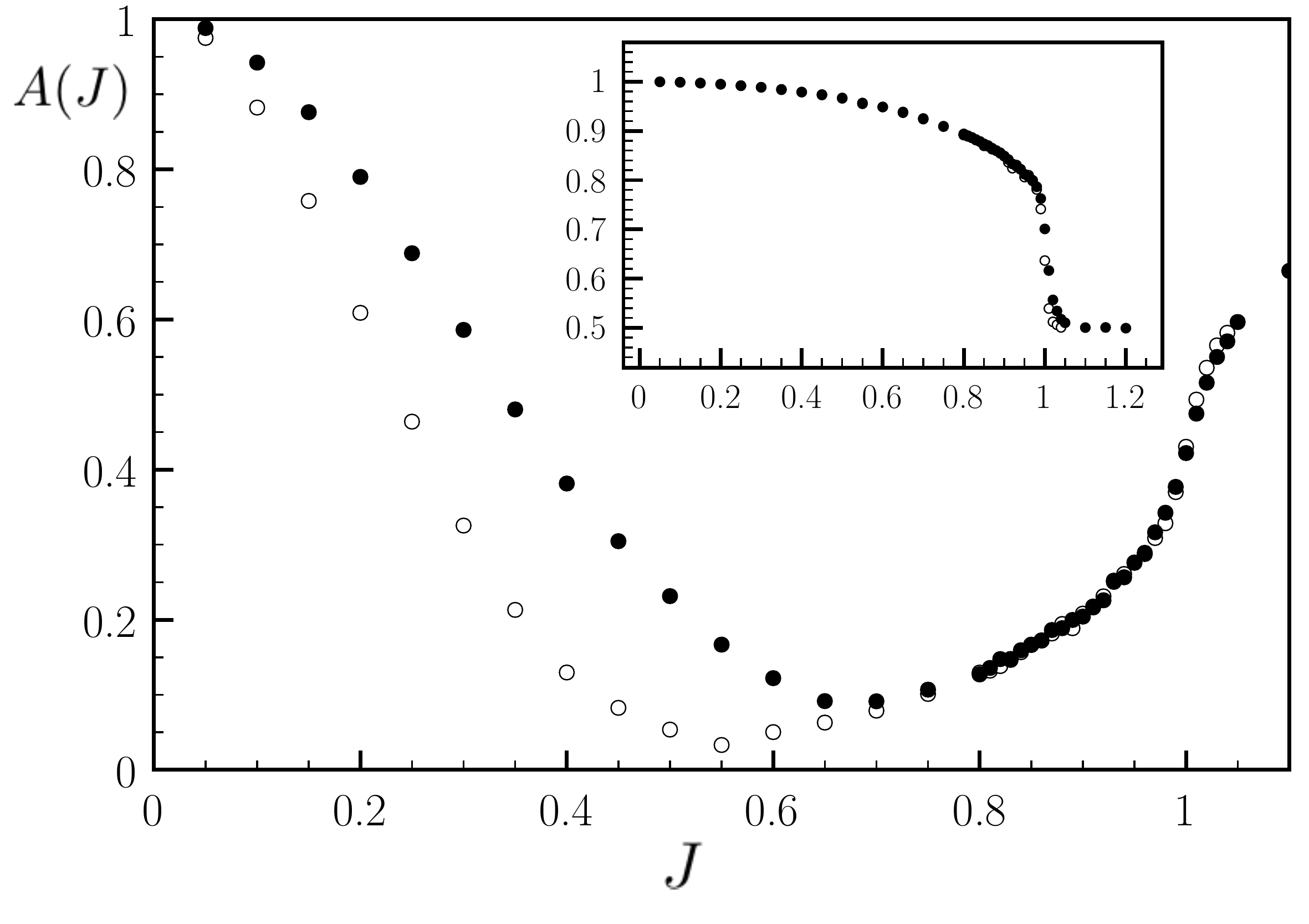}
    \caption{The amplitude $A(J)$ for the sinusoidal behavior of $\expt{xz}$
    after a projection $P$ and $SO(2)$ correction on one qubit is done on the state.
    Closed circles are for $N=50$ while open circles are $N=102$. The inset shows
    the probability $p_{yy}$ of success for the measurement. Note that in the
    latter case the two system sizes are practically indistinguishable.}
    \label{Fig:AJ}
\end{figure}

We can now investigate the behaviour of our localized entangled
state as we approach the phase transition at $J=1$.  We restrict our
sampled states to maximally-entangled ones by projecting onto the
subspace of states $\ket\Phi$, i.e. using a projector
$P=\ket{C_{00}}\bra{C_{00}}+\ket{C_{11}}\bra{C_{11}}$.  The
probability of this projection viewed as a measurement is denoted
$p_{yy}$; we note that this probability is large for all $J$ up to
the phase transition as seen in the inset of Fig.~\ref{Fig:AJ}.

\begin{figure}
    \includegraphics[width=\columnwidth]{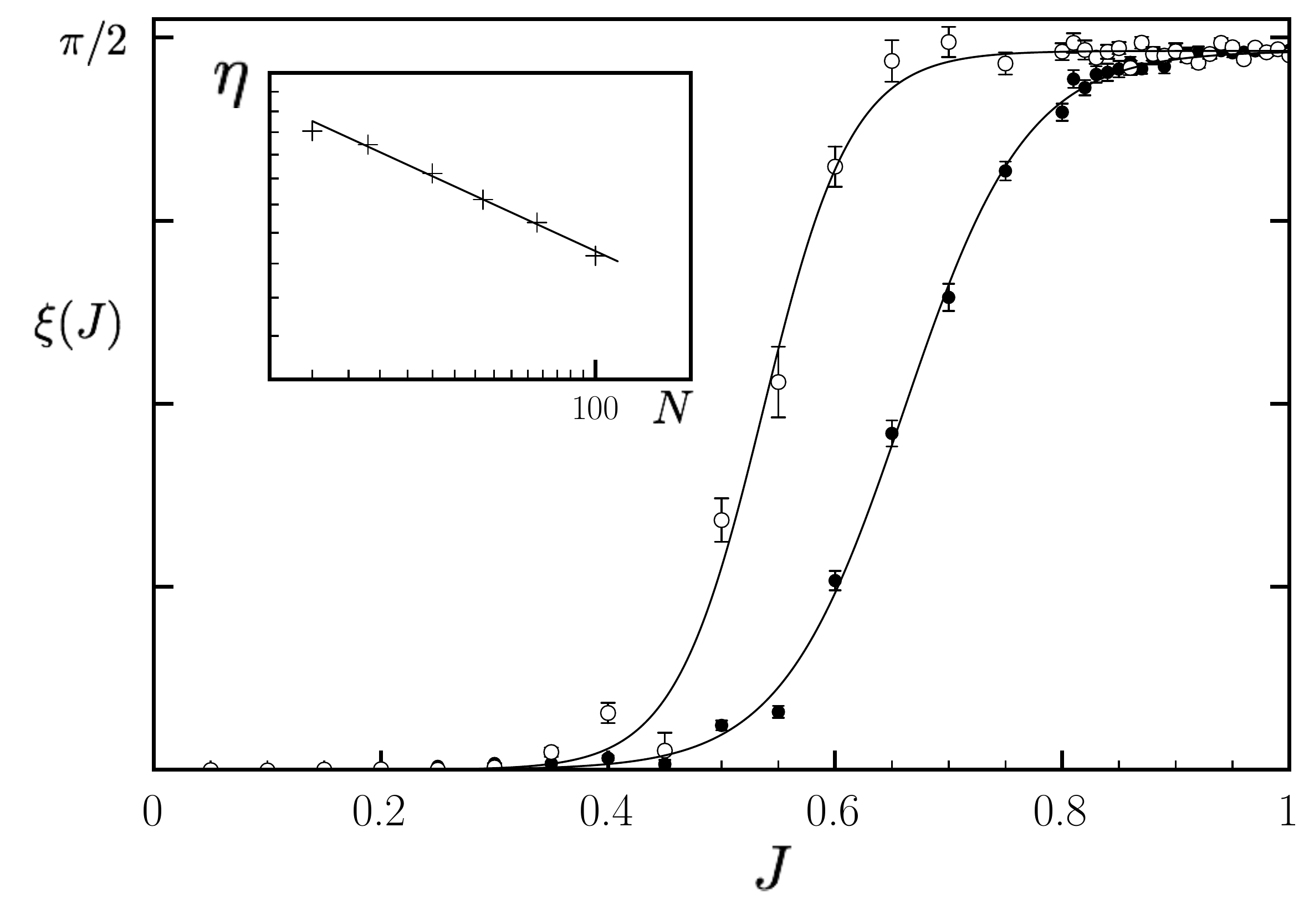}
    \caption{The phase $\xi(J)$ is shown for $N=50$ (closed circles) and $N=100$
    (open circles). The error bars refer to the fitting to the sinusoidal curve.
    Full lines are best fits to $\tanh$, while the inset shows the scaling law
    for the transition parameter $\eta\sim N^{-1/3}$ on double log scale.}
    \label{Fig:tanh}
\end{figure}

On states following the projection, we investigate the average
angle.  We find that the phase $\xi(J)$ is an indicator of a
size-dependent transition, changing rapidly from zero to $\pi/2$,
with a behavior closely approximated by
\begin{equation}
   \xi(J)=\frac\pi4\left(\tanh\{K(N)[J-\eta(N)]\}+1\right)\,,
\end{equation}
where the coefficients $K$ and $\eta$ are size dependent. The
parameter $\eta$ is a good marker for the transition, and as shown
in Fig.~\ref{Fig:tanh}, it follows a power law $\eta\sim N^{-1/3}$,
while $A$ increases non-universally with $N$.  Up to this
transition, there is no bias, and the state approximates a cluster
state.  Beyond this transition, the distribution $\mathcal
D_J(\Phi)$ is almost uniform, meaning that there is no rotation
independent of the measurement results that will increase the
fidelity with the two-qubit cluster state.

In summary, while the localizable EL remains infinite throughout the
cluster phase, the resulting two-qubit entangled state cannot be
deterministically transformed to the fiducial two-qubit cluster
state using Pauli corrections for any $J>0$.  Further work is
required to determine the non-Pauli correction angle (about the
$y$-axis) based on the measurement results.  Potentially, this
non-Pauli rotation could be useful for developing quantum
computational wires~\cite{GrossEisert:2008} that perform non-trivial
single-qubit gates.  However, without characterizing this rotation
in terms of the measurement results, such states cannot be used
directly as quantum computational wires.

\section{Disentangling measurements}
\label{Sec:Disentangle}

Finally, in addition to localizable entanglement, we consider
another property of quantum states that is useful (though possibly
not necessary) for measurement-based quantum computation.  The
cluster state possesses the useful property that a $Z$ measurement
on any qubit ``removes'' that qubit and leaves the remaining qubits
in a cluster state (up to a Pauli correction dependent on the
measurement result).  On a 1-D chain, a $Z$ measurement on a qubit
will then \emph{disentangle} the two halfs of the remaining chain.
This property is shared by our model Hamiltonian for any $B_z$ as
long as $B_x=J=0$~\cite{Barrett:2008}.  If we turn instead to the
model with $\vec{B}=0$ but $J>0$, we find that on an $N=3$ open
chain, any measurement on the middle qubit will result in some
remaining entanglement on the two end qubits, and therefore be an
inadequate disentangling measurement.

\begin{figure}
    \includegraphics[width=\columnwidth]{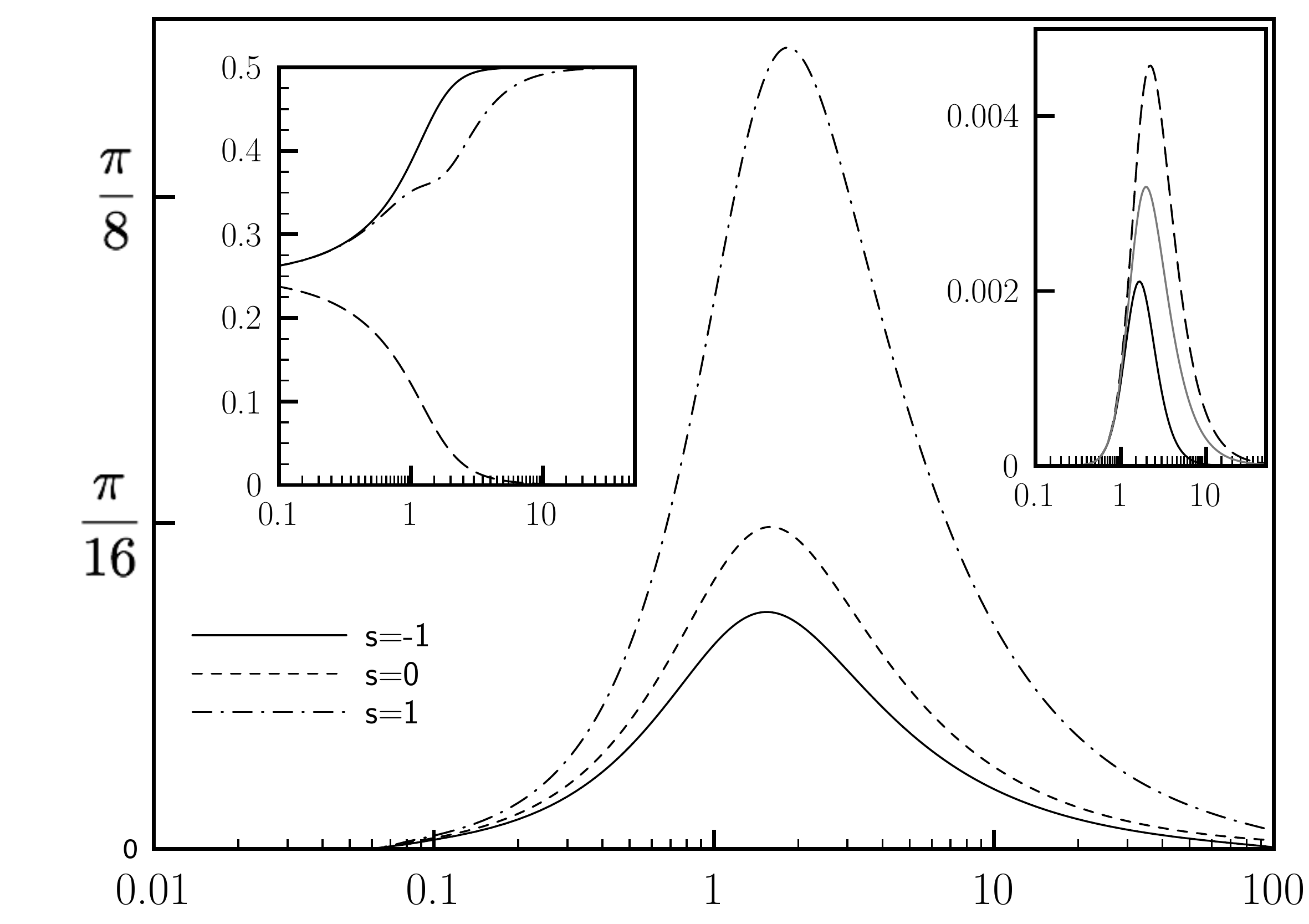}
    \caption{For a chain of 4 qubits, the measurement angle that disentangles the
     two extreme qubits is shown as a function of $J$ in the main panel. The
     left inset shows the corresponding outcome probabilities for the three possible
     measurement outcomes. The right inset shows the remaining entanglement
     multiplied with the probability of the outcome between the end qubits if no
     correction is done. In this case the outcomes $s=\pm1$ are identical, so there
     are only two distinct cases, and the average entanglement is shown as a grey
     line.}
    \label{Fig:Disentangler}
\end{figure}

In this section, we consider using local measurements on
\emph{pairs} of qubits that optimally disentangle the remaining
halves of a 1-D chain with $\vec{B}=0$.  (To be clear, we still
consider \emph{single}-qubit measurements, applied to two
neighbouring qubits.)  First, we consider both qubits to be measured
at an angle $\vartheta$ in the $x{-}z$ plane. The outcomes $s$ are
labelled according to the total spin in the desired direction which
is either of $s=\{-1, 0, 1\}$. There are two outcomes with total
spin zero, but these lead to identical states due to the reflection
symmetry of the model and are therefore treated together. The
optimal measurement angle depends on the measurement outcome, and
this dependency is shown in Fig. \ref{Fig:Disentangler}.
Because the optimal measurement angle depends on the measurement
outcome. Note that while all four measurement outcomes (there are
two corresponding to $s=0$) are equally probable in the cluster
state, the probability of the $s=0$ result vanishes for the GHZ
state while the two remaining are equally probable.

Alternatively, one can view the measurement as a weak measurement
described by four POVMs described in terms of the three measurement
angles, $\vartheta_p$ with $p$ the measurement outcome, as shown in
Fig.~\ref{Fig:Disentangler}. The POVMs are thus
\begin{equation}
    \begin{split}
    E_{-1}&=cM_-(\vartheta_{-1})\otimes M_-(\vartheta_{-1})\\
    E_0   &=cM_-(\vartheta_{0})\otimes M_+(\vartheta_{0})\\
    E'_0  &=cM_+(\vartheta_{0})\otimes M_-(\vartheta_{0})\\
    E_1   &=cM_+(\vartheta_{1})\otimes M_+(\vartheta_{1})\\
    E_X   &=\mathds 1-E_{-1}-E_0-E'_0-E_1,
    \label{POVMdef}
    \end{split}
\end{equation}
where $M_\pm(\vartheta)$ is a local projective measurement along the
direction $\vartheta$, and the constant approximately $c\leq.5$ is
required for $E_X$ to be positive in general. The element $E_X$,
consisting of all local measurement outcomes that do not correspond
to one of the others, is considered a `failure' outcome.  If this
measurement result is obtained, the result must be discarded, while
for any other result, the measurement sequence disentangles the two
outer qubits.  The probability of failure is $p(E_X)=\bra\psi
E_X\ket\psi$. For the specific model in question, $c\leq.8$ is
sufficient to retain positivity for all $J$. Since a higher $c$
means a lower probability of failure, on can further minimize
failure by selecting the highest $c$ that allows for a positive
$E_X$ for a specific set of measurement angles.

One may imagine that the POVM can be replaced by a projective
two-qubit measurement, or equivalently that one can make adaptive
measurements without postselection. However, this is not possible
for any $0<J<\infty$. Consider making two measurements in the four
qubit chain with different angles, $\theta_1$ and $\theta_2$.
Mapping out the combinations that disentangle the chain for all four
measurement outcomes, a pure projective measurement would be
possible if all four lines crossed at some combination, while an
adaptive or entangled measurement if one (say) $\theta_1$ for an
outcome meant one could pick out a definite $\theta_2$ depending on
the outcome and be ensured disentanglement. However, our numerical
investigation has determined that this situation does not occur for
any finite $J$.  Hence, postselection is the only way to ensure
disentanglement.

Note that these results for the $N=4$ chain cannot be immediately
extrapolated to a general $N$ qubit chain, but the same general
results hold, and a disentangling measurement can still be
constructed by equal measurements on the $(N-2)/2$ middle qubit
pairs with disentangling angles close to those described here.

\section{Conclusions}
\label{Sec:Conclusions}

For a 1D spin chain for which the ground state is a cluster state,
we have demonstrated the existence of a robust ``cluster phase''
under local ($B_x$-field) and coupled (Ising) perturbations.  All
states in this phase exhibit diverging localizable EL.  However, as
we have been unable to determine the bi-product unitary as a
function of the measurement results, the usefulness of such states
as a quantum computational wire may be limited.

The existence of this cluster phase in one dimension may be extended
to two dimensions even though our current numerical procedures are
unlikely to be suitable for this. When extending to higher
dimensions by similar schemes to MPS~\cite{Banuls:2008}, the
entanglement properties are likely to be even harder to distill.
Alternative techniques beyond the MPS paradigm may be more
suitable~\cite{Vidal:2007}.

\begin{acknowledgments}
  SOS is funded by the Research Council of Norway.
  SDB acknowledges the support of the Australian Research Council.
  We thank Andrew Doherty and Terry Rudolph for helpful discussions.
\end{acknowledgments}

\appendix

\section{Jordan-Wigner transform}
\label{Appendix:JW}

The Jordan Wigner transform \cite{JordanWigner} is a mapping from
the spins defined by the Pauli operators into spinless fermions, and
is a standard technique in condensed matter physics. We give a quick
overview here, and show how to obtain the relevant correlation
functions. For a $N$ qubit chain, $N$ spinless fermions are defined
by annihilation operators
\[a_\mu=\frac12\left(\prod_{\nu<\mu}\sigma^z_\nu\right)(\sigma^x_\mu+i\sigma^y_\mu)\]
and the adjoint creation operators. Futher one defines $2N$ self
adjoint Majorana fermions
\[\gamma_{2\mu-1}=\frac1{i\sqrt2}\left(a_\mu-a^\dag_\mu\right)\quad
\gamma_{2\mu}=\frac1{\sqrt 2}\left(a_\mu+a^\dag_\mu\right)\] where
$\{\gamma_i,\gamma_j\}=\delta_{ij}$.  Switching coordinates
$x\leftrightarrow z$ for the Hamiltonian (\ref{Eq:Hamiltonian}) with
$B_z=0$ gives
\[H=\sum_{ij}\gamma_i\mathcal C_{ij}\gamma_j\]
with the $2N\times 2N$ matrix
\[\mathcal C=i\begin{pmatrix}
\mathcal B&\mathcal J&\mathcal I&0&\cdots&P\mathcal I\transpose&P\mathcal J\transpose\\
-\mathcal J\transpose&\mathcal B&\mathcal J&\mathcal I&\cdots&0&P\mathcal I\transpose\\
-\mathcal I\transpose&-\mathcal J\transpose&\mathcal B&\mathcal J&&\cdots&0\\
0&-\mathcal I\transpose&-\mathcal J\transpose&\mathcal B&&\cdots&0\\
&&\vdots\\
-P\mathcal J&-P\mathcal I&0&\cdots&-\mathcal I\transpose&-\mathcal J\transpose&\mathcal B
\end{pmatrix}\]
where
\[\mathcal B=\begin{pmatrix}0&-B_x\\B_x&0\end{pmatrix}\quad
\mathcal J=\begin{pmatrix}0&J\\0&0\end{pmatrix}\quad
\mathcal I=\begin{pmatrix}0&1\\0&0\end{pmatrix}
\]
and $P=\prod\sigma^z_\mu=\pm1$ is the parity of the model which is a
conserved quantity, $[H,P]=0$. If one considers a chain with open
boundary conditions this simply amounts to setting $P=0$. It is not
immediately obvious which parity segment the ground state belongs
to, but this is easy to verify by computing the two energies. For
large $N$ the difference between the two obtained states are in any
respect very small.

Define the correlation matrix
$\Gamma_{ij}=\expt{[\gamma_i,\gamma_j]}$, in which every second
entry is zero since
$\expt{\gamma_{2\mu}\gamma_{2\nu}}=\expt{\gamma_{2\mu+1}\gamma_{2\nu+1}}=0$.
This enables us to compute the von Neumann entropy of any subset of
qubits effectively \cite{Latorre:2004p1285}, and expectation values
also follow easily.

The local $z$ expectation value (which corresponds to $x$ in the
original coordinates) is simply
$\expt{\sigma^z_\mu}=\Gamma_{2\mu-1,2\mu}=\Gamma_{12}$ where the
last step is valid under PBC, or $P\not=0$. As an example of a more
complicated expectation value, consider (\ref{ZXexptvalue}), which
in the current coordinates and transformed to Majorana fermions is
\begin{align*}
    \mathcal O&=\expt{\sigma^x_2\sigma^z_3\sigma^z_5\cdots\sigma^z_{N-1}\sigma^x_N}\\
    &=\expt{\gamma_3\left(\prod_{k=1}^{N/2-2}\gamma_{4k+3}\gamma_{4(k+1)}\right)\gamma_{2N}}.
\end{align*}
Using Wick's theorem, and the fact that only odd/even combinations
of indices couples, one finds that $\mathcal O=\det G$ where $G$ is
a submatrix of $\Gamma$,
\[G=\begin{pmatrix}
\Gamma_{3,8}&\Gamma_{3,12}&\Gamma_{3,16}&\cdots&\Gamma_{3,2N-4}&\Gamma_{3,2N}\\
\Gamma_{7,8}&\Gamma_{7,12}&\Gamma_{7,16}&\cdots&\Gamma_{7,2N-4}&\Gamma_{7,2N}\\
\Gamma_{11,8}&&\cdots\\
\vdots\\
\Gamma_{2N-5,8}&\Gamma_{2N-5,12}&&\cdots&&\Gamma_{2N-5,2N}
\end{pmatrix}.\]
Indeed, any expectation value is the determinant of a dense
submatrix of $\Gamma$, the size of which depends on the expectation
value in question.

\end{document}